\documentclass{josr}

\usepackage[english]{babel}
\usepackage[utf8]{inputenc}
\usepackage{listings}
\usepackage{url}
\usepackage{xcolor}
\usepackage{booktabs}

\usepackage{amsmath}

\usepackage{doi}
\usepackage{graphicx}
\usepackage{listings}
\colorlet{punct}{red!60!black}
\definecolor{background}{HTML}{EEEEEE}
\definecolor{delim}{RGB}{20,105,176}
\colorlet{numb}{magenta!60!black}

\lstdefinelanguage{json}{
    basicstyle=\normalfont\ttfamily,
    numbers=left,
    numberstyle=\scriptsize,
    stepnumber=1,
    numbersep=8pt,
    showstringspaces=false,
    breaklines=true,
    frame=lines,
    backgroundcolor=\color{background},
    literate=
     *{0}{{{\color{numb}0}}}{1}
      {1}{{{\color{numb}1}}}{1}
      {2}{{{\color{numb}2}}}{1}
      {3}{{{\color{numb}3}}}{1}
      {4}{{{\color{numb}4}}}{1}
      {5}{{{\color{numb}5}}}{1}
      {6}{{{\color{numb}6}}}{1}
      {7}{{{\color{numb}7}}}{1}
      {8}{{{\color{numb}8}}}{1}
      {9}{{{\color{numb}9}}}{1}
      {:}{{{\color{punct}{:}}}}{1}
      {,}{{{\color{punct}{,}}}}{1}
      {\{}{{{\color{delim}{\{}}}}{1}
      {\}}{{{\color{delim}{\}}}}}{1}
      {[}{{{\color{delim}{[}}}}{1}
      {]}{{{\color{delim}{]}}}}{1},
}

\usepackage[
    backend=biber,
    style=numeric,
    natbib=true,
    sortlocale=en_US,
    url=true,
    doi=true,
    isbn=false,
    eprint=false,
    sorting=none,
]{biblatex}
\renewcommand{\vec}[1]{\boldsymbol{#1}}

\usepackage{pifont}% http://ctan.org/pkg/pifont
\newcommand{\cmark}{\ding{51}}%
\newcommand{\xmark}{\ding{55}}%

\newcommand{\n}{Glyph$\;$}
\newcommand{\repo}{\emph{\url{https://github.com/Ambrosys/glyph}}}
\newcommand{\mytitle}{Glyph: Symbolic Regression Tools}

\definecolor{mygray}{gray}{0.6}

\usepackage{hyperref}

\begin{document}

{\bf Software paper for submission to the Journal of Open Research Software} \\

To complete this template, please replace the blue text with your own. The paper has three main sections: (1) Overview; (2) Availability; (3) Reuse potential. \\

Please submit the completed paper to: editor.jors@ubiquitypress.com

\rule{\textwidth}{1pt}

\section*{(1) Overview}

\vspace{0.5cm}

\section*{Title}
\mytitle

\section*{Paper Authors}
1. Quade, Markus, markus.quade@ambrosys.de \\
2. Gout, Julien, julien.gout@ambrosys.de \\
3. Abel, Markus, markus.abel@ambrosys.de

\section*{Paper Author Roles and Affiliations}

1. Quade, Markus, lead developer, Ambrosys GmbH, David Gilly Stra\ss{}e 1, Potsdam, Germany \\
2. Gout, Julien, core contributer, Ambrosys GmbH, David Gilly Stra\ss{}e 1, Potsdam, Germany \\
3. Abel, Markus, scientific project lead, Ambrosys GmbH, David Gilly Stra\ss{}e 1, Potsdam, Germany

\section*{Abstract}

We present Glyph - a Python package for genetic programming based symbolic regression. \n is designed for usage let by numerical simulations let by real world experiments. For experimentalists, glyph-remote provides a separation of tasks: a ZeroMQ interface splits the genetic programming optimization task from the evaluation of an experimental (or numerical) run. \n can be accessed at \repo.  Domain experts are be able to employ symbolic regression in their experiments with ease, even if they are not expert programmers. The reuse potential is kept high by a generic interface design. Glyph is available on PyPI and Github.

\section*{Keywords}

Symbolic Regression; Genetic Programming; MLC; Python

\section*{Introduction}

Symbolic regression \cite{schmidt2009distilling} is an optimization method to find an optimal representation of a function. The method is ``symbolic'', because building blocks of the functions, i.e. variables, primitive functions, and operators, are represented symbolically on the computer. Genetic programming (GP) \cite{koza1992genetic} can be implemented to find such a function for system identification \cite{Vladislavleva2013,Quade2016} or fluid dynamical control \cite{gout2016,MLC}. \n is an effort to separate optimization method and optimization task allowing domain-experts without special programming skills to employ symbolic regression in their experiments. We adopt this separation of concerns implementing a client-server architecture; a minimal communication protocol eases its use. Throughout this paper ``experiment'' is meant as a synonym for any symbolic regression task including a lab-experiment, a numerical simulation or data fitting.

Previous work on system identification and reverse engineering of conservation laws was  reported in \cite{schmidt2009distilling,schmidt2011automated}. Modern algorithms also include multi objective optimization \cite{Quade2016} and  advances like age fitness based genetic programming \cite{Schmidt2011} or epigenetic local search \cite{LaCava2016}. There exist various approaches to the representation of multi IO problems, including stack- or graph-based representations and pointers \cite{LaCava2016,Galvan-Lopez2008}.

\section*{Implementation and architecture}

\n is intended as a lightweight framework to build an application finding an optimal system representation given measurement data. The main application is intended as system control, consequently a control law is determined and returned. \n is built on the idea of loose coupling such that dependencies can be released if wanted.

\begin{figure}[h]
    \includegraphics[width=0.5\linewidth]{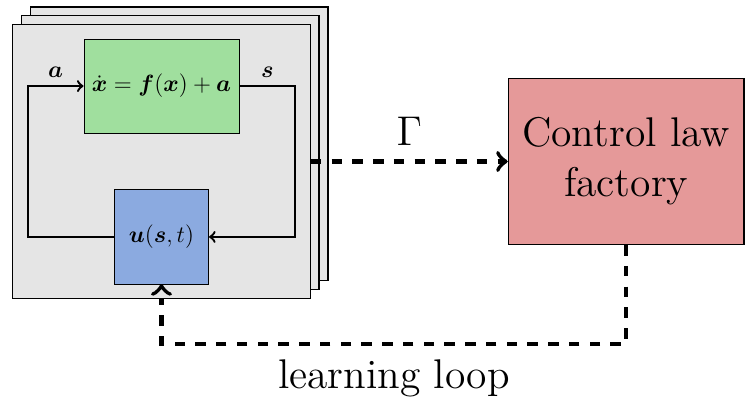}
    \centering
    \caption{Left: A typical closed loop control task is sketched. Given a system $\dot{\vec{x}} = \vec{f}(\vec{x})$, some measurements $s$ and a control law $\vec{u}(\vec{s},t)$ and we can control the system by adding the actuation $\vec{a} = \vec{u}(\vec{s}, t)$. Right: gp-based symbolic regression finds different candidate control laws. Each candidate solution is given a fitness score $\Gamma$ which is used to compare different solutions and to advance the search in function space. Figure adapted from \cite{gout2016} with permission.}
    \label{fig:learning_loop}
\end{figure}

A typical control application consists of a system and its controller, possibly separated, cf. Fig.~\ref{fig:learning_loop}.
\n has three main abstractions to build such an application: i) the assessment, which holds all methods and data structures belonging to the experiment, ii) the GP which is responsible for the system identification, and iii) the application components, which constitute an application.

% mention that we focus here on the application rather on the library
\subsection*{Building an Application}

An application consists of a GP callable, the gp\_runner, an assessment callable for input, the assessment\_runner, and the application which uses both of these classes and holds all application-relevant details. A command-line application is built by
\begin{lstlisting}[language=python,firstnumber=1]
    assessment_runner = AssessmentRunner(assess_args)
    gp_runner = glyph.application.GPRunner(gp_args)
    app = command_line_application(app_args)
\end{lstlisting}

The assessment\_runner has one argument, the parallel\_factory which implements a map() method, possibly parallel. For an application one needs to implement setup, assign\_fitness, and measure: setup is self-explaining, measure is a key method which takes as input a set of measurement functions and combines them into a tuple of callable measures for multiobjective optimization. The measures are used eventually in assign\_fitness where the return values are used to assign a fitness to an individual from GP. The interface is freely extensible. A gp\_runner forwards the evolutionary iteration. It takes as arguments, gp\_args, an individual class, a gp\_algorithm, and an assessment\_runner.
The individual class contains the representation of a function, the individual; it is currently based on deap's tree-based implementation. The gp\_algorithm takes care for the breeding and selection steps, its principles are described in \cite{koza1992genetic}.

The application is run in the main function with \lstinline{app.run()}. Each of the high-level functions contains a bunch of next-level instructions, and can be built with a minimal assembly of methods.

In the application and gp\_runner, the user has freedom to add functionality using the list of callbacks in the arguments, say, to implement other logging or streaming options. This allows for very flexible programming. We constructed the components that way to allow users to specialize for their particular experiments and possibly increase performance or extend the symbolic regression, e.g. by replacing the deap tree-based representation of an individual.

\subsection*{Remote Control}
\begin{figure}[h]
    \includegraphics[width=0.6\linewidth]{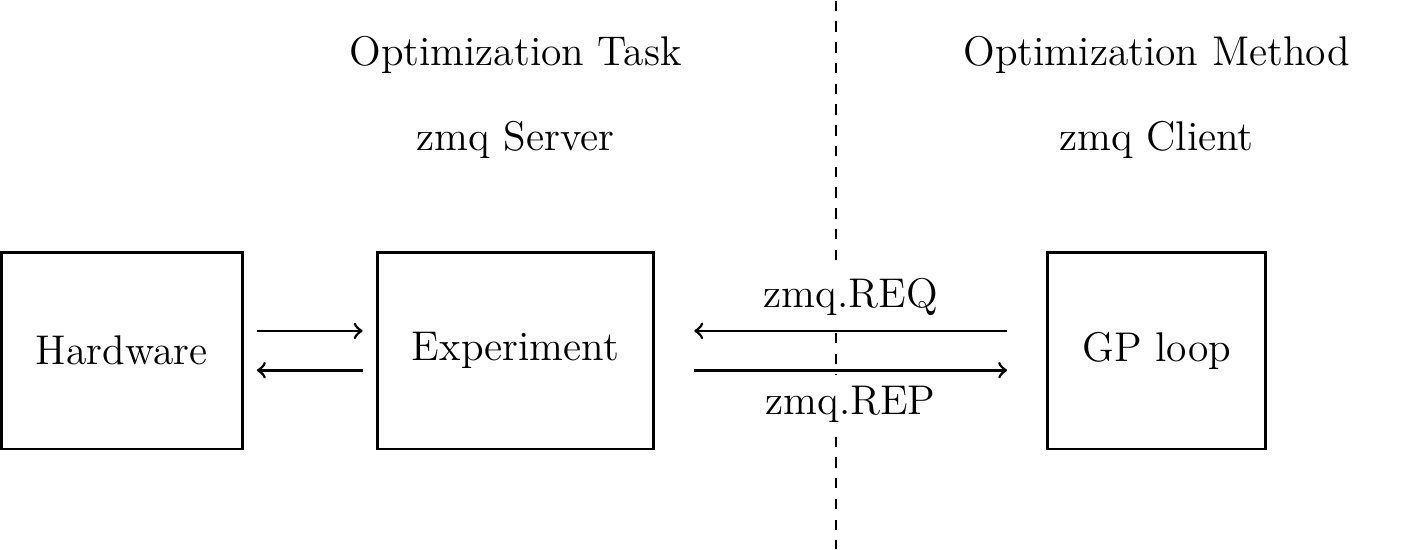}
    \centering
    \caption{Sketch of the implementation of the experiment - GP communication as client-server pattern. Left: single experiment server plus event handler. Right: GP client. Both parts are interfaced using ZeroMQ. As described in Sec.~\ref{sec:cprotocol} the GP program performs requests, e.g. the evaluation of a candidate solution. The event handler takes care of these requests and eventually forwards them to the hardware.}
    \label{fig:communication}
\end{figure}

One main objective of glyph is its use in a real experiment. In this case, the GP loop is separated from the experimental loop in a client-server setup using ZeroMQ \cite{Akgul:2013:ZER:2523409}, cf. Fig.~\ref{fig:communication}.

Consequently, one should implement the interface to the experiment using the protocol described in Sec.~\ref{sec:cprotocol}. Having the implementation of the experiment, the server, one needs to implement the client, i.e. the interface to the gp\_runner. In essence this means connecting the correct sockets whith ZeroMQ and ensuring that the gp\_runner and the assessment\_runner use the corresponding sockets. Then, the main application is assembled as before, now using a \textit{RemoteApp} for the main application, which in turn uses a gp\_runner, which then uses now a RemoteAssessmentRunner. That is it, we can run remotely our GP evaluation from some client and the experiment in place of the experiment.

\subsection*{Communication Protocol\label{sec:cprotocol}}
The communication is encoded in json \cite{ecma404}. A message is a json object with two members:
\begin{lstlisting}[language=json,firstnumber=1]
{
    "action": "value",
    "payload": "value",
}
\end{lstlisting}

The possible values are listed in Table~\ref{tab:json}.
\begin{table}
\centering
\begin{tabular}{l|l|l|l}
Action name & Payload & Expected return Value \\ \hline
\emph{CONFIG} & -- & config settings  \\
\emph{EXPERIMENT} & list of expressions & list of fitness value(s) \\
\emph{SHUTDOWN} & -- & --\\
\end{tabular}
\caption{Communication protocol.}
\label{tab:json}
\end{table}
The config action is performed prior to the evolutionary loop. Entering the loop, discovered solutions will be batched and a \emph{experiment} action will be requested. You can configure optional caching for re-discovered solutions. This includes persistent caching between different runs. The \emph{shutdown} action will let the experiment program know that the gp loop is finished and you can safely stop the hardware.

Configuration settings are sent as a json object in key:value form, where the keys contain the option to be set, there is only one mandatory option: the primitive set.
To configure the primitive set, the primitive names are passed as content of the key \emph{config}, whose values specify the corresponding arities, both fields described
again as json object.

The \emph{experiment} action sends a list of expressions, encoded as string in prefix (also: polish) notation \cite{jorke1989arithmetische}. For each expression sent, the experiment returns a fitness tuple.

Additionally, one can define the type of algorithm, error metric, representation, hyperparameters, etc. A comprehensive up to date list can be found at \url{http://glyph.readthedocs.io/en/latest/usr/glyph_remote/}.

\subsection*{Application example: control of the chaotic Lorenz System}\
In the following, we demonstrate the application and use of \n by the determination of an unknown optimal control law for a chaotic system.
As an example, we study the control of the potentially chaotic Lorenz system. Chaotic systems are very hard to predict and control in practice due to their sensitivity towards small changes in the initial state which may lead to exponential divergence of trajectories. The Lorenz model \cite{Lorenz1963} consists of a system of three ordinary differential equations:
\begin{align}
\begin{split}
    \dot{x} &= s (y - x)\label{eqn:lorenz}\\
    \dot{y} &= r x - y - x z \\
    \dot{z} &= x y - b z,
\end{split}
\end{align}
with two nonlinearities, $xy$ and $xz$. Here $x$, $y$, and $z$ make up the system state and $s$, $r$, $b$ are parameters: $s$ is the Prandtl number, $r$ is the Rayleigh
number, and b is related to the aspect ratio of the air rolls. For a certain choice of parameters and initial conditions chaotic behavior emerges.

Here we present two examples where the target is to learn control of bring a chaotic Lorenz system to a complete stop, that is, $(x,y,z) = 0$ ($t \in I \!\!R$). In the first example, the actuator term is applied to $\dot{y}$. This allows for a more direct control of the system, since $y$ appears in every equation of \eqref{eqn:lorenz} and, thus, influence all three state components, $x$, $y$, and $z$. In the second example the actuator term is applied to $\dot{z}$, which leads to a more indirect control, since the flow of information from $z$ to $x$ is only through $y$.

\begin{table}[h]
    \caption{General setup of the GP runs. \label{tab:setup}}
    \centering
    \begin{tabular}{ll}
        \toprule
        \addlinespace
        population size & $500$ \\
        max. generations & $20$\\
        MOO algorithm & NSGA-II \\
        \addlinespace
        tree generation & \textit{halfandhalf} \\
        min. height & 1 \\
        max. height & 4 \\
        \addlinespace
        selection & \textit{selTournament} \\
        tournament size & $2$ \\
        \addlinespace
        breeding & \textit{varOr} \\
        \addlinespace
        recombination & \textit{cxOnePoint} \\
        crossover probability & $0.5$ \\
        crossover max. height & $20$ \\
        \addlinespace
        mutation & \textit{mutUniform} \\
        mutation probability & $0.2$ \\
        mutation max. height & $20$ \\
        \addlinespace
        constant optimization & \textit{leastsq} \\
        \addlinespace
        \bottomrule
    \end{tabular}
\end{table}

\begin{table}[htpb]
    \caption{Control of the Lorenz system: system setup.\label{tab:lorenz_y_setup}}
    \centering
    \begin{tabular}{lllll}
        \toprule
        \multicolumn{2}{c}{dynamic system} & & \multicolumn{2}{c}{GP} \\
        \cmidrule{1-2} \cmidrule{4-5}
        \addlinespace
        $s$         & $10$      && cost functionals & RMSE$(x,0)$ \\
        $r$         & $28$      &&                  & RMSE$(y,0)$ \\
        $b$         & $8/3$     &&                  & RMSE$(z,0)$ \\
        $x(t_0)$    & $10.0$    &&                  & length$(u)$ \\
        $y(t_0)$    & $1.0$     && argument set     & $\{x, y, z\}$ \\
        $z(t_0)$    & $5.0$     && constant set     & $\{k\}$ \\
        $t_0,\,t_n$ & $0,\,100$ && seed (in $y$)    & $4360036820278701581$ \\
        $n$         & $5000$    && seed (in $z$)    & $2480329230996732981$ \\
        \addlinespace
        \bottomrule
    \end{tabular}
\end{table}

\begin{figure}[htpb]
    \centering
    \includegraphics[width=\textwidth]{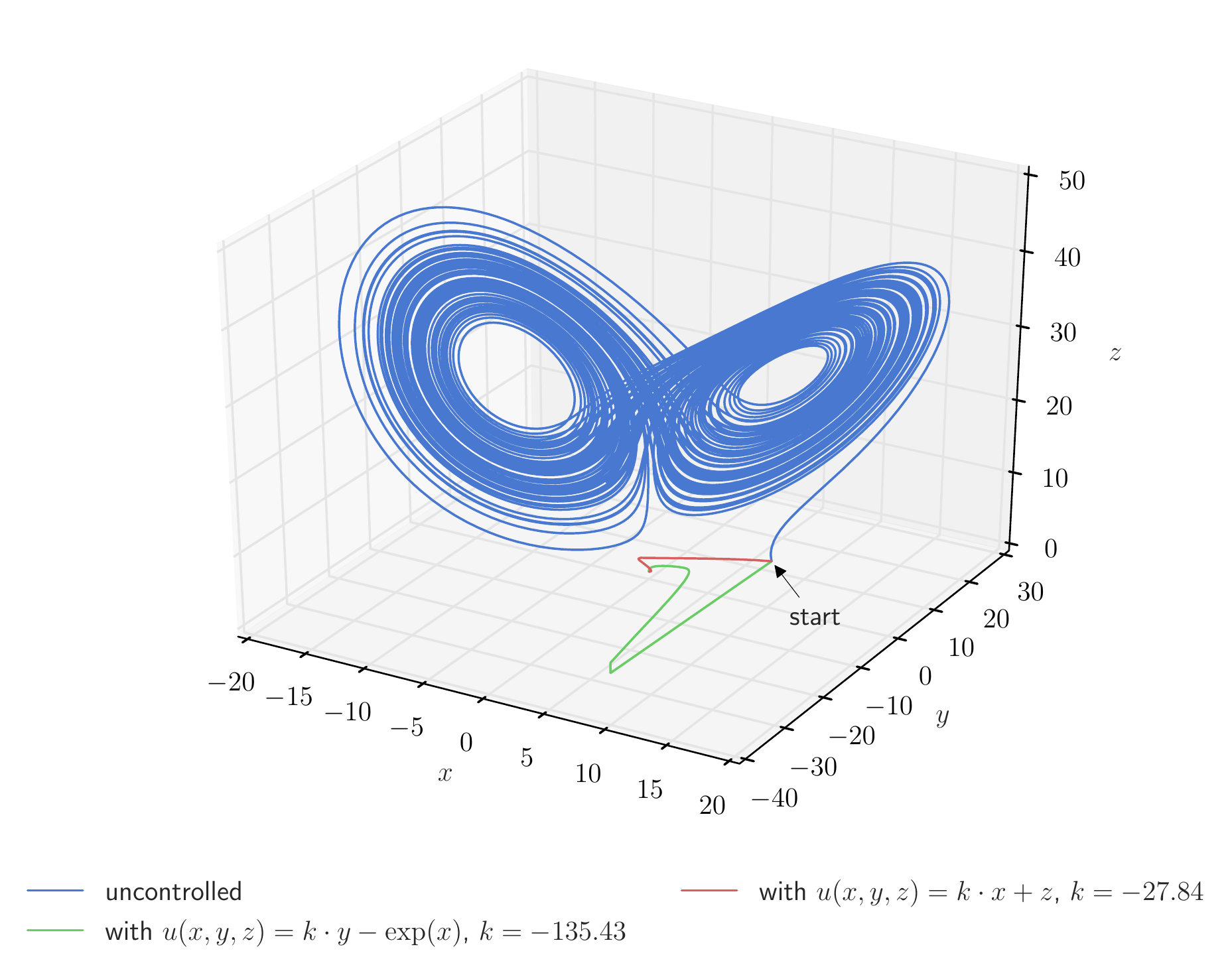} \\
    \caption{Phase portrait of the forced Lorenz system with control exerted in $\dot{y}$. (Green and red: The system trajectories when controlled by two particular Pareto-front solutions. Blue: the uncontrolled chaotic system.)\label{fig:lorenz_y_phaseportrait}}
\end{figure}

The system setup is summarized in Table~\ref{tab:setup} and Table~\ref{tab:lorenz_y_setup}. When $r = 28$, $s =10$, and $b =8/3$, the Lorenz system produces chaotic solutions (not all solutions are chaotic). Almost all initial points will tend to an invariant set -- the Lorenz attractor -- a strange attractor and a fractal. When plotted the chaotic trajectory of the Lorenz system resembles a butterfly (blue graph in Fig.~\ref{fig:lorenz_y_phaseportrait}). The target of control is, again, formulated as RMSE of the system state with respect to zero (separately for each component)
\begin{align*}
\begin{split}
\Gamma_1 := \text{RMSE}(x, 0),\;
\Gamma_2 := \text{RMSE}(y, 0),\;
\Gamma_3 := \text{RMSE}(z, 0).
\end{split}
\end{align*}

The control function $u$ can make use of ideal measurements of the state components. Constant optimization is performed on a single constant $k$. The respective GP runs for control in $y$ and control in $z$ are conducted with the corresponding random seeds labeled ``in $y$'' and ``in $z$''.

\paragraph{Control in $\boldsymbol{y}$:}
For control in $y$ the actuator term $u$ is added to the left side of the equation for $\dot{y}$ in the uncontrolled system \eqref{eqn:lorenz}:
\begin{align*}
\begin{split}
    \dot{y} &= r x - y - x z + u(x, y, z).
\end{split}
\end{align*}

The Pareto solutions from the GP run are shown in Table~\ref{tab:lorenz_y_results}. The wide spread of the cost indices is a sign of conflicting objectives that are hard to satisfy in conjunction. Interestingly, almost all solutions, $u$, commonly introduce a negative growth rate into $\dot{y}$. This effectively drives $y$ to zero and suppresses the growth terms, $sy$ and $xy$, in the equations for $\dot{x}$ and $\dot{z}$ respectively, in turn, driving $x$ and $z$ to zero as well. As would be expected, minimal expressions, of length 1 or 2, cannot compete in terms of the RMSE. For example, the simple solution, $u(x,y,z)=-ky$ (fourth row), is almost as good as the lengthier one, $u(x,y,z)=-\exp(x) + ky$ (first row), and even better in RMSE$_y$.

\begin{table}[htpb]
    \caption{Control of the Lorentz system in $y$: Pareto-front solutions.\label{tab:lorenz_y_results}}
    \centering
    \begin{tabular}{ccccll}
        \toprule
        RMSE$_x$ & RMSE$_y$ & RMSE$_z$ & length & {expression} & {constants} \\
        \midrule
        \addlinespace
        $0.178884$ & $0.087476$ & $0.105256$  & $ 7$ & $-\exp(x) + k \cdot y                $ & $k = -135.43$ \\
        $0.241226$ & $0.069896$ & $0.213063$  & $ 5$ & $k \cdot x + z                       $ & $k = -27.84$ \\
        $0.246315$ & $0.014142$ & $0.222345$  & $ 6$ & $- z + k \cdot y                     $ & $k = -75590.65$ \\
        $0.246316$ & $0.014142$ & $0.222347$  & $ 4$ & $-k \cdot y                          $ & $k = 75608.50$ \\
        $0.246367$ & $0.028851$ & $0.220426$  & $10$ & $-x \cdot (k + y) \cdot \exp(\exp(y))$ & $k = 9.62$ \\
        $0.246729$ & $0.118439$ & $0.211212$  & $ 6$ & $-x \cdot (k + y)                    $ & $k = 29.21$ \\
        $0.246850$ & $0.031747$ & $0.220726$  & $ 9$ & $-x \cdot (k + y) \cdot \exp(y)      $ & $k = 26.12$ \\
        $4.476902$ & $4.468534$ & $7.488516$  & $ 3$ & $-\exp(y)                            $ & \\
        $7.783655$ & $8.820086$ & $24.122441$ & $ 2$ & $-x                                  $ & \\
        $7.931978$ & $9.066296$ & $25.047630$ & $ 1$ & $k                                   $ & $k = 1.0$ \\
        $8.319191$ & $8.371462$ & $25.932887$ & $ 2$ & $-y                                  $ & \\
        $8.994685$ & $9.042226$ & $30.300641$ & $ 1$ & $z                                   $ & \\
        \addlinespace
        \bottomrule
    \end{tabular}
\end{table}

Table~\ref{tab:lorenz_y_results} shows the results from the GP run. One solution immediately stands out: $u = k \cdot x + z$, with $k = -27.84$ (second row). It is exactly what one might expect as a control term for the chaotic Lorenz system with control in $y$.  This control law effectively reduces the Rayleigh number $r$ to a value close to zero ($k \approx r$), pushing the Lorenz system past the first pitchfork bifurcation, at $r = 1$, back into the stable-origin regime. If $r < 1$ then there is only one equilibrium point, which is at the origin. This point corresponds to no convection. All orbits converge to the origin, which is a global attractor, when $r < 1$.

\begin{figure}[htpb]
    \centering
    \includegraphics[width=.31\textwidth]{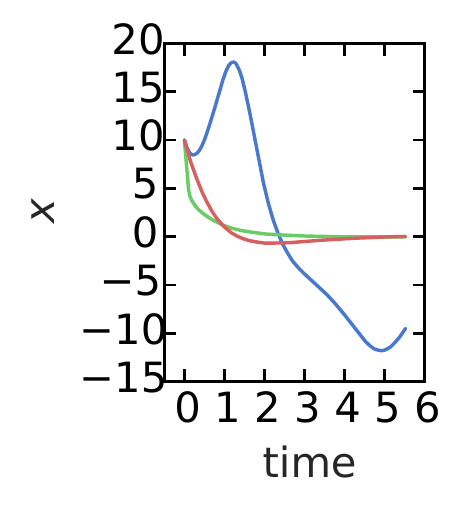}
    \includegraphics[width=.31\textwidth]{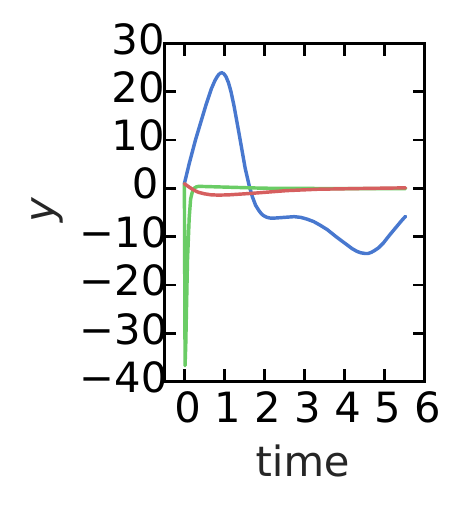}
    \includegraphics[width=.31\textwidth]{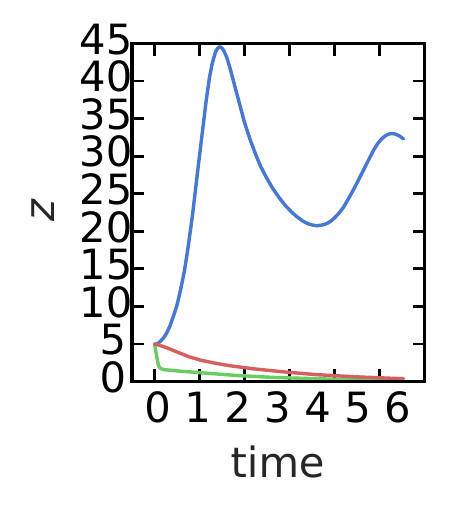}
    \caption{Detailed view of the single trajectories in $x$, $y$, and $z$ dimension. (blue: uncontrolled; green: $u(x,y,z) = -\exp(x) + k \cdot y$, $k=-135.43$; red: $u(x,y,z) = k \cdot x + z$, $k=-27.84$.) \label{fig:lorenz_y_xyz}}
\end{figure}

The phase portrait of the solution from the first and second row of Table~\ref{tab:lorenz_y_results} are illustrated in Fig.~\ref{fig:lorenz_y_phaseportrait}. After a short excursion in negative $y$ direction ($t \approx 5$), the green trajectory quickly converges to zero. The red trajectory seems to take a shorter path in phase space, but, it is actually slower to converge to the origin. This is verified by a plot of the trajectories for the separate dimensions $x$, $y$ and $z$ over time Fig.~\ref{fig:lorenz_y_xyz}.

\paragraph{Control in $\boldsymbol{z}$:}

For control in $z$ the actuator term $u$ is added to the left side of the equation for $\dot{z}$ in the uncontrolled system \eqref{eqn:lorenz}
\begin{align*}
\begin{split}
    \dot{z} &= x y - b z + u(x, y, z) \\
\end{split}
\end{align*}

\begin{figure}[htpb]
    \centering
    \includegraphics[width=1.\textwidth]{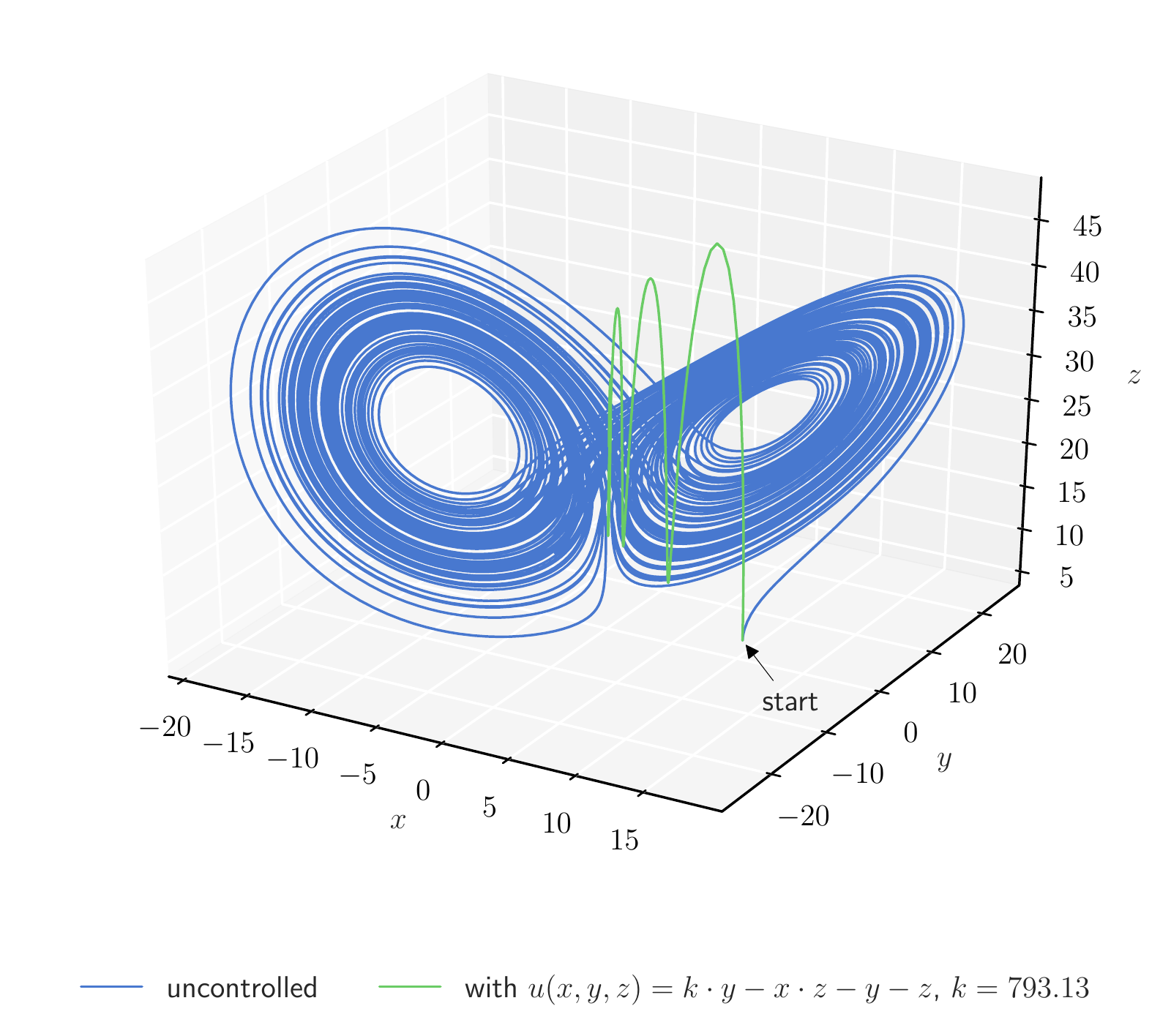} \\
    \caption{Control of the Lorentz system in $\dot{z}$.\label{fig:lorenz_z_phaseportrait}}
\end{figure}

Selected Pareto-front individuals from the GP run are displayed in Table~\ref{tab:lorenz_z_results}. As mentioned at the beginning of this section, effective control is hindered by the indirect influence of $z$ on the other state variables, hence, it is not surprising that the control laws here are more involved than in the previous case. Also, they generally do not perform well in the control of $z$, which is expressed by the relatively high values in RMSE$_z$. This is confirmed by the phase portrait of the solution $u(x,y,z)=- (k \cdot (-y) + x \cdot z + y + z)$ shown in figure Fig.~\ref{fig:lorenz_z_phaseportrait}: While going straight to the origin in the $xy$-plane there are strong oscillations of the trajectory along the $z$-axis.

The dynamics caused by the actuation, e.g. for the best control law found, can be explained qualitatively: there is a strong damping in all variables but $y$. This reflects the tendency to suppress $z$-oscillations and, at the same time, to add damping in $y$ through the $xz$ term: if $y$ grows, the $z$ contribution to damping on the right hand side of the Lorenz equations \eqref{eqn:lorenz} grows and, in turn, damps $y$. This is, however, only possible to some extent, hence, the oscillations observed in figure Fig.~\ref{fig:lorenz_z_phaseportrait}.

\begin{table}[htpb]
    \caption{Control of the Lorentz system in $z$: selected Pareto-front solutions.\label{tab:lorenz_z_results}}
    \centering
    \begin{tabular}{ccccll}
        \toprule
        RMSE$_x$ & RMSE$_y$ & RMSE$_z$ & length & {expression} & {constants} \\
        \midrule
        \addlinespace
        $0.289289$ & $0.139652$ & $26.994070$ & $13$ & $- (k \cdot (-y) + x \cdot z + y + z)      $ & $k = 793.129676$ \\
        $0.327926$ & $0.267043$ & $27.070289$ & $ 8$ & $\exp(-k + y \cdot \sin(y))                $ & $k = -4.254574$ \\
        $0.431993$ & $0.508829$ & $32.116326$ & $ 7$ & $(k + x) \cdot (y + z)                     $ & $k = 2.638069$ \\
        $0.471535$ & $0.525010$ & $26.986321$ & $ 5$ & $k + x \cdot z                             $ & $k = 67.137183$ \\
        $0.637056$ & $0.605686$ & $26.895493$ & $ 7$ & $\exp(k + y \cdot \sin(y))                 $ & $k = 3.964478$ \\
        $0.677204$ & $0.703577$ & $27.019308$ & $ 4$ & $y + \exp(k)                               $ & $k = 4.276256$ \\
        $0.930668$ & $0.952734$ & $26.895126$ & $ 5$ & $x + \exp(\exp(k))                         $ & $k = 1.448198$ \\
        $1.764030$ & $1.860288$ & $26.766383$ & $ 6$ & $(k + x) \cdot \exp(y)                     $ & $k = 21.783557$ \\
        \addlinespace
        \bottomrule
    \end{tabular}
\end{table}

We conclude the demonstration with a short summary: Using \n we can find complex control laws, even for unknown systems. This cannot be easily achieved with other frameworks. The control laws found can be studied analytically in contrast to several other methods which have black-box character. The  usage is straightforward, as we have described above. The above example can be found online as an example.

\subsubsection*{Other symbolic regression libraries}

Due to its popularity, symbolic regression is implemented by most genetic programming libraries. A semi-curated list can be found at \url{http://geneticprogramming.com/software/}.
In contrast to other implementations, \n implements higher concepts, such as symbolic constant optimization, and also offers parallel execution for complex examples (control simulation, system identification). \n is well tested, cf. Table~\ref{tab:comparison} and currently applied in two experiments and several numerical problems.
For control, there exists a dedicated matlab toolbox (with python interface), openMLC \cite{machinelearningcontrol_2017}, which contains much of the material treated in \cite{MLC}.

\begin{table}
\begin{tabular}{|l|c|c|c|c|c|c|c|c|}
\hline
        & CI/tests & doc & open api & caching & checkpointing & MOGP & SCO & MO\\
\hline
\hline
openMLC & \cmark & \xmark & \xmark & \cmark & \cmark & \xmark & \xmark & \xmark\\
\hline
Glyph & \cmark & \cmark & \cmark & \cmark & \cmark & \cmark& \cmark & \cmark \\
\hline
\end{tabular}
\centering
\caption{Comparison of \n and openMLC features. MOGP refers to multi objective optimization. MO means multiple outputs. SCO means symbolic constant optimization.}
\label{tab:comparison}
\end{table}

\section*{Quality control}

Continuous Integration tests are conducted for Mac, Linux and Windows using Travis and AppVoyer. The tests consider Python version 3.5 and 3.6 . Unit test coverage is around 85\% as reported by codecov. Additionally, tests specifically cover the stochastic parts of the optimization to ensure reproducibility. Along with the software tests are shipped which guarantee the correct execution of the examples. The user can reuse these tests for further development. Locally, tests can be executed via the pytest command.

\section*{(2) Availability}
\vspace{0.5cm}
\section*{Operating system}

\n is compatible with Mac, Linux and Windows.

\section*{Programming language}

Python 3.5+

\section*{Dependencies}

Currently, \n is based on \emph{DEAP} \cite{DEAP_JMLR2012}, an evolutionary computation framework adopting a toolbox-like structure for rapid prototyping.  Further dependencies are found up-to-date at \url{https://github.com/Ambrosys/glyph/blob/master/requirements.txt}.

\section*{List of contributors}

Core contributors (prior to open source):
Markus Quade
Julien Gout

Open source contributors can be found at \url{https://github.com/Ambrosys/glyph/graphs/contributors}.

\section*{Software location:}

{\bf Archive} Zenodo

\begin{description}[noitemsep,topsep=0pt]
	\item[Name:] Ambrosys/glyph
	\item[Persistent identifier:] \url{http://doi.org/10.5281/zenodo.801819}
	\item[Licence:] LGPL
	\item[Publisher:]  Markus Quade
	\item[Version published:] 0.3.3
	\item[Date published:] 12.09.17
\end{description}

{\bf Code repository} Github

\begin{description}[noitemsep,topsep=0pt]
	\item[Name:] glyph
	\item[Persistent identifier:] \url{https://github.com/Ambrosys/glyph}
	\item[Licence:] LGPL
	\item[Date published:] 08.12.16
\end{description}

\section*{Language}

English

\section*{(3) Reuse potential}

The potential to use \n is twofold: one one hand applications can be easily written and the elegant core functionality can be extended; on the other hand, researchers can use the code as core for symbolic regression and extend its functionality in a very generic way. With respect to applications, currently two main directions are targeted: modeling using genetic programming- based symbolic regression and the control of complex system, where a control law can be found generically, using genetic programming. The detailed examples and tutorial allow usage from beginner to experienced level, i.e. undergraduage research projects to faculty research. The design of \n is such that generic interfaces are provided allowing for very flexible extension.

\section*{Acknowledgements}

We acknowledge very fruitful discussions with respect to applications of MLC with S. Brunton,  B. Noack, A. Pikovsky, M. Rosenblum, R. Semaan, and B. Strom and coding gossip with V. Mittal.

\section*{Funding statement}

This work has been partially supported by the German Science Foundation via SFB 880. MQ was supported by a fellowship within the FITweltweit program of the German Academic Exchange Service (DAAD).

\section*{Competing interests}

The authors declare that they have no competing interests.

\printbibliography

% \bibliography{main}{}
% \bibliographystyle{plainurl}

\end{document}